\documentclass[aps,twocolumn,pre,showpacs,eqsecnum]{revtex4-1}
\usepackage{graphpap}
\usepackage[dvips]{graphicx}
\usepackage[dvips]{graphics}
\usepackage{color}

\usepackage[normalem]{ulem} 
\usepackage {ulem} 

\begin{document}

\title{Etched graphene quantum dots on hexagonal boron nitride}
 \author{S. Engels$^{1,2}$, A. Epping$^{1,2}$, C. Volk$^{1,2}$, S. Korte$^{2}$, B. Voigtl\"ander$^{2}$, K.~Watanabe$^3$, T. Taniguchi$^3$, 
 S. Trellenkamp$^{2}$, 
and C. Stampfer$^{1,2}$}
\affiliation{
$^1$JARA-FIT and II. Institute of Physics, RWTH Aachen University, 52074 Aachen, Germany \\
$^2$Peter Gr\"unberg Institute (PGI-3/9), Forschungszentrum J\"ulich, 52425 J\"ulich, Germany \\
$^3$National Institute for Materials Science, 1-1 Namiki, Tsukuba, 305-0044, Japan \\
}
\date{ \today}

\begin{abstract}
We report on the fabrication and characterization of etched graphene quantum dots (QDs) on hexagonal boron nitride (hBN) and SiO$_2$ with different island diameters.
We perform a statistical analysis of Coulomb peak spacings over a wide energy range. For graphene QDs on hBN, the standard deviation of the normalized peak spacing distribution decreases with increasing QD diameter, whereas for QDs on SiO$_2$ no diameter dependency is observed. In addition, QDs on hBN are more stable under the influence of perpendicular magnetic fields up to 9T. Both results indicate a substantially reduced substrate induced disorder potential in graphene QDs on hBN.
\end{abstract}

 \maketitle

\newpage

Graphene promises weak spin-orbit~\cite{hue06,min06} and hyperfine interaction \cite{tra07} making this material interesting for hosting quantum dots (QDs) with potentially long-living spin states. However, the missing band gap in graphene makes the confinement of electrons challenging. 
At present, there are two main strategies to overcome this limitation: (i) size confinement by nanostructuring
 \cite{han07, tod09, sta09, mol09, liu09, sta08a, pon08, sch09, mor09, liu10,cai10} or, (ii) top-gating bilayer graphene \cite{cas07, oos07, tay10, per07, all12, goo12}.
In the first case, the broken lattice symmetry introduces an effective energy gap, while in bilayer graphene a transverse electric field breaks the inversion symmetry resulting in a small band gap~\cite{oos07}. 
This second approach has only recently been demonstrated to be
promising for confining carriers~\cite{all12, goo12}. Vice versa, nanostructured graphene QDs have been
 extensively studied in the last years and e.g. electron-hole crossover~\cite{gue09}, spin states~\cite{gue10} and charge relaxation times~\cite{vol13} have been reported. However, graphene nanodevices suffer from disorder, making it hard to tune QDs into the few
 carrier regime. The disorder potential in these devices is expected to arise both from the substrate and the edge roughness. A promising approach to reduce the substrate (i.e. bulk) disorder is based on placing graphene on hexagonal boron nitride (hBN) \cite{dea10,xue11}.
  While graphene on SiO$_2$ exhibits charge puddles with diameters on the order of a few tens of nm~\cite{mar08}, the size of charge puddles in graphene on hBN have been reported to be roughly one order of magnitude larger \cite{xue11}. These results make hBN an interesting substrate also for graphene QDs, and it may allow to learn more about the contribution of edge roughness to the overall disorder.

\begin{figure}[t]\centering
\includegraphics[draft=false,keepaspectratio=true,clip,%
         width=0.95\linewidth]%
                   {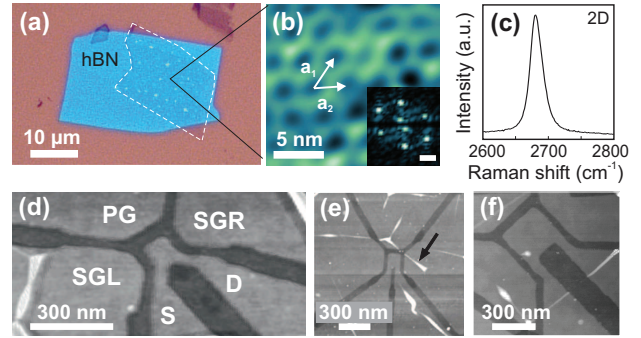}
\caption[fig01]{(color online) (a) Optical image of a graphene flake on hBN. (b) Fourier filtered STM image of graphene on hBN exhibiting a periodic Moir\'e pattern. Inset: Fourier spectrum of the unfiltered data in the main panel. Scale bar is 0.5~nm$^{-1}$. (c) 2D peak of a representative Raman spectrum of a  transfered graphene flake on hBN. (d)-(f) Scanning force micrographs of etched graphene quantum dots on hBN with different diameters ((d) $d$=110~nm, (e) $d$=180~nm and (f) $d$=300~nm).}
\label{fig01}
\end{figure}

In this letter, we investigate nanostructured graphene QDs on hBN with island diameters ranging from 100 to 300~nm. These values are on the order of the expected size of charge puddles in bulk graphene on hBN. To allow for a comparative study, we fabricated geometrically identical graphene QDs on 
SiO$_2$ and performed low temperature transport measurements on both kind of devices. In particular, we focus on the fluctuations of the Coulomb-peak spacings as a function of the dot size. We show that, for graphene QDs on hBN, the standard deviation of the normalized peak spacing distribution decreases with increasing island diameter.  Vice versa, for QDs on SiO$_2$ no diameter dependence can be observed in the investigated regime. In addition, we show that QDs on hBN exhibit a stable single-dot behavior even in magnetic fields up to 9T. All results indicate that the disorder potential is significantly reduced in graphene QDs on hBN with larger diameter and that edge contribution dominates the disorder potential for smaller QDs.

The device fabrication is based on mechanical exfoliation of graphene and hBN flakes.
Hexagonal BN flakes are deposited on 295~nm SiO$_2$ on highly doped Si substrates.
By closely following the work of Dean et al.~\cite{dea10}, we transfered individual graphene
sheets on selected hBN flakes with a thickness of around 20-30~nm (see example in Fig.~1(a)).
Electron beam lithography (EBL) followed by reactive ion etching with an Ar/O$_2$ plasma is employed to etch the graphene flakes.
The resulting graphene nanostructures are then contacted in a second EBL step, followed by metal evaporation of Cr/Au (5~nm/150~nm). 

To verify the single-layer nature of the transfered graphitic films we perform Raman spectroscopy measurements. In Fig.~1(c) we show the 2D peak of a typical Raman spectrum. The 2D peak is centered at 2680~cm$^{-1}$ and exhibits a full width at half maximum (FWHM) of 24~cm$^{-1}$, which proves that the investigated flake is a single-layer graphene sheet. Our fabrication process for graphene on hBN has been optimized to obtain high quality samples with low doping fluctuations and a low overall doping level, as detailed in Ref.~\cite{for13}.
Fig.~1(b) shows a scanning tunneling microscope (STM) image of 
a graphene flake on hBN.
Measurements are performed in a multi-tip STM setup at a bias voltage of 0.5~V and a constant current of 65~pA.
The data are Fourier filtered around the high symmetry points of the unfiltered two-dimensional Fourier spectrum (bright spots in the inset of Fig.~1(b)) to enhance the visibility of the emergent periodic pattern. This can be identified as a Moir\'{e} pattern arising from the lattice mismatch of graphene and hBN, reflecting the high quality of the transfered graphene. The unit cell vectors $a_1$ and $a_2$ have a length of about 3~nm and are consistent with a Moir\'{e} pattern arising from an angular lattice mismatch of less than 5$^{\circ}$ at the investigated location~\cite{yan12}. 

In Figs.~1(d)-(f), we show scanning force microscope (SFM) images of etched graphene quantum dot devices on hBN with different island sizes. All devices are based on a graphene island that is connected by constrictions to both source (S) and drain (D) leads. Lateral graphene gates (side gate left (SGL), PG, side date right (SGR) in Fig.~1(d)) are placed nearby the island and the constrictions, and allow to locally tune the chemical potential. All devices are intentionally located in the center of areas which are not disturbed by the characteristic wrinkles of graphene on hBN (see e.g. arrow in Fig.~1(e)). To allow for a detailed comparative study, we fabricated identical graphene QDs on SiO$_2$. In all devices, the underlying highly doped Si substrate can be used as a back gate (BG) to adjust the Fermi level.

\begin{figure}[tb]\centering
\includegraphics[draft=false,keepaspectratio=true,clip,%
                   width=1.0\linewidth]%
                   {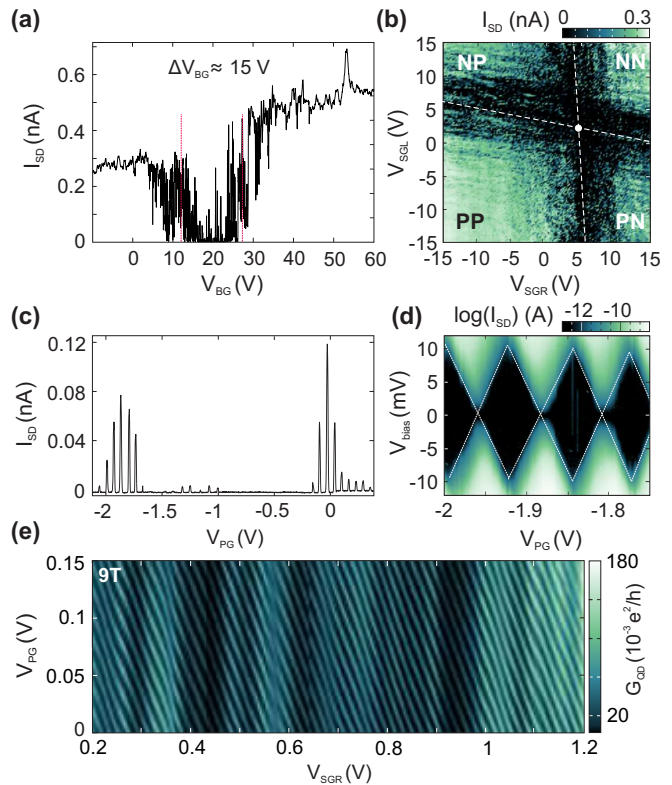}
\caption[fig02]{(color online) (a) Back gate ($V_{\rm{BG}}$) dependence of the current $I_{SD}$ through the graphene QD shown in Fig.~1(d) ($d=110$~nm) at a constant $V_{\rm{bias}}$. A transport gap of $\Delta V_{\rm{BG}} \approx15$~V around $V_{\rm{BG}}=18$~V is visible. (b) $I_{\rm{SD}}$ as function of the side gate voltages $V_{\rm{SGR}}$ and $V_{\rm{SGL}}$ at $V_{\rm{BG}}$=22~V and $V_{\rm{bias}}$=300~$\mu$V. The left and right graphene constrictions can be separately tuned into the hole (P) and electron transport regime (N). (c) Plunger gate ($V_{\rm PG}$) dependence of $I_{\rm{SD}}$ at constant $V_{\rm{SGR}}$=5.2~V and $V_{\rm{SGL}}$=2.4~V showing sharp Coulomb peaks. (d) Finite bias spectroscopy measurements exhibiting Coulomb diamonds. (e) Charge stability diagram of a graphene QD on hBN with a diameter of $d$=300~nm at $B$=9~T ($V_{\rm{bias}}$=300~$\mu$V).}
\label{fig02}
\end{figure}

In Figs.~2(a)-(d), we show low-temperature transport measurements (1.5~K) performed on a QD device on hBN, with an island diameter of 110~nm (see Fig. 1(d)).
Fig.~\ref{fig02}(a) shows the source-drain current as function of back gate voltage, $V_{\rm{BG}}$ (side gate voltages are at 0~V). 
The so-called transport gap, i.e. the region of suppressed current around $V_{BG}$ = 20~V extends over a range of roughly 
$\Delta V_{\rm{BG}} \approx$ 15~V, which is 
in agreement with earlier studies on etched graphene nanoribbons and QDs
  on SiO$_2$~\cite{han07, tod09, sta09, mol09, liu09, sta08a, pon08, sch09, mor09, liu10,cai10} and graphene nanoribbons on hBN~\cite{bis12}.
In Fig.~\ref{fig02}(b), the current $I_{SD}$ is recorded as a function of the side gate voltages $V_{\rm{SGR}}$ and $V_{\rm{SGL}}$. The cross shaped region of suppressed current can be attributed to the transport gap of the two constrictions connecting the island to the leads.
Similarly to the device discussed in Ref.~\cite{sta08a}, the small cross-talk of the lateral side gates allow 
to tune transport through the two constrictions independently into the electron (N) and hole (P) regime.
The resulting 4 different regimes with finite current (NN, NP, PP and PN)  are indicated in Fig.~2(b).
In our study we used this type of maps for fixing the side gate voltages deep in the common transport gap (see dot and dashed lines in Fig.~2(b)), where
both constrictions perform best as tunneling barriers.  
In Fig.~\ref{fig02}(c), we show the current $I_{\rm{SD}}$ as function of the plunger gate (PG) voltage $V_{\rm PG}$ in a regime where the two barriers are pinched-off. Distinct resonances occur due to Coulomb blockade in the graphene QD. From finite bias measurements (see Fig.~2(d)) we extract a PG lever arm of $\alpha_{} \approx$ 0.15 and a charging energy of $E_C \approx 8-10$~meV, which is in reasonable agreement with values reported earlier for graphene QDs of similar size on SiO$_2$~\cite{sch09,vol13}. 
In Fig.~2(e) we show the charge stability diagram of a graphene QD on hBN with a diameter $d$ = 300~nm, in the presence of a perpendicular magnetic field of 9~T. The vertical features in the measurement can be attributed to resonances located in the right constriction, while the diagonal lines of elevated conductance correspond to the Coulomb peaks in the actual QD. This measurement indicates that, even at this high magnetic fields, the sample behaves as a single QD over a very large range of gate voltages. Such a high stability of single-dot characteristics is hard to observe in graphene QDs of similar size on SiO$_2$, which tend to break apart into several smaller dots in high magnetic fields due to the roughness of the disorder potential~\cite{gue11b}. This observation is a first indication of a rather homogeneous potential landscape for graphene QDs on hBN, compared to SiO$_2$.

\begin{figure}[tb]\centering
\includegraphics[draft=false,keepaspectratio=true,clip,%
                   width=1.0\linewidth]%
                   {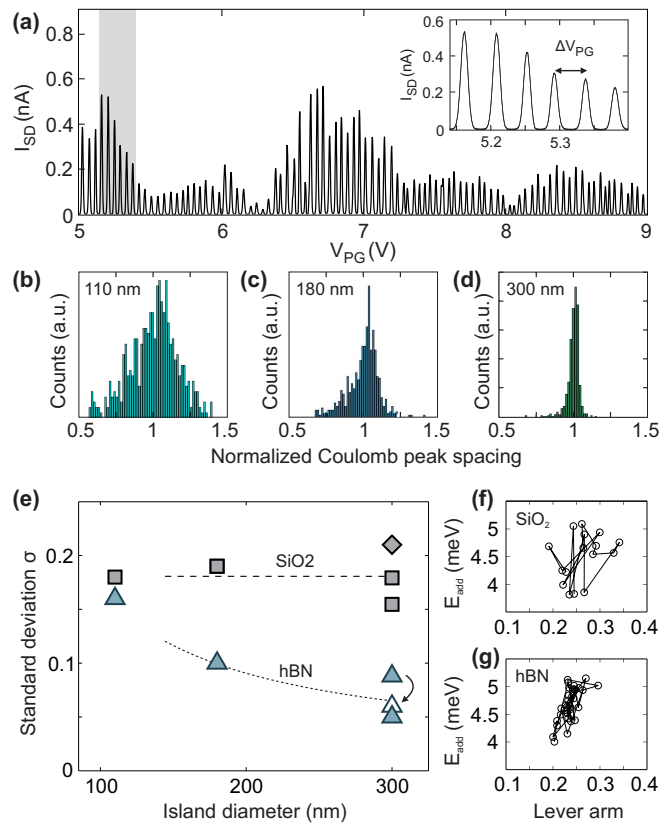}
\caption[fig03]{(color online) (a) Source-drain current $I_{\rm{SD}}$ as a function of of $V_{\rm PG}$ for a QD on hBN with $d$=180~nm. The inset shows a close-up  of the shaded region. (b)-(d) Normalized peak-spacing distribution of QDs on hBN with diameters of (b) $d$=110~nm,(c) $d$=180~nm and (d) $d$=300~nm. (e) Summary plot of the standard deviation $\sigma$ of the normalized peak-spacing distribution for different QD sizes on a SiO$_2$ (rectangular data points) and hBN (triangular data points) substrate. One of the two QDs with $d$=300~nm has been measured first at $B$=0~T and then at $B$=9~T (see arrow and white triangular). (f) and (g) Dependence of the PG lever arm on the addition energy $E_{\rm add}$ for QDs with $d$=300~nm on SiO$_2$ (f) and hBN (g).}
\label{fig03}
\end{figure}

For a more detailed and quantitative comparison between graphene QDs resting on hBN and SiO$_2$
we study in total 8 different devices fabricated on both substrates, and focus in particular on the 
distribution of the Coulomb-peak spacing $\Delta V_{\rm PG}$, i.e. the spacing  between two subsequent Coulomb peaks (see inset in Fig.~\ref{fig03}(a)).

A typical series of 94 Coulomb peaks of a QD on hBN with $d$ = 180~nm is shown in Fig.~\ref{fig03}(a). The spacings between two consecutive peaks show no systematic tendency towards lower or higher values with varying gate voltage.
Similar measurements are performed for various graphene QDs with diameters ranging from $d$=110~nm to 300~nm, and around 600 Coulomb peaks are analyzed for each device~\cite{numberofPeaks}. The observed normalized Coulomb peak spacings $\Delta V_{\rm PG}/\overline{\Delta V_{\rm PG}}$ for QDs on hBN are reported as histograms in Figs.~\ref{fig03}(b)-(d), where $\overline{\Delta V_{\rm PG}}$ is the mean peak spacing of each device. The distribution of peak spacings shows a clear narrowing for larger island sizes. More quantitatively, the standard deviation of the normalized peak-spacing distribution reads 0.16 for the QD with $d=$ 110~nm, and it decreases to 0.10 and 0.05 for the dots with $d=$ 180~nm and $d=$ 300~nm, respectively.

The same kind of measurements are performed also on geometrically identical QDs on SiO$_{2}$. A summary of these results is given in Fig.~\ref{fig03}(e), where we plot the standard deviation of the normalized peak spacing distribution for all measured QDs as a function of the island size. Each filled data point corresponds to a different device (the diamond-shaped one is obtained from the  earlier measurements discussed in Ref. \cite{sta08a}). A striking difference can be observed between QDs on hBN and geometrically identical devices on SiO$_{2}$. While in the first case, the standard deviation of the peak-spacing distribution shows a clear dependence on the island diameter $d$, in the second case it is independent of $d$, within fluctuations between devices.
In Figs.~3(f)-(g), we show the dependency of  the addition energy $E_{\rm add}=E_{C}+\Delta$ on the plunger gate lever arm $\alpha$ for QDs on SiO$_2$ (Fig. 3(f)) and hBN (Fig. 3(g)) with a diameter of $d=$ 300~nm. Here, $\Delta$ is the single-particle level spacing. The data are extracted from up to 30 subsequent Coulomb diamonds, similar to those of Fig.~\ref{fig02}(d), and show that the fluctuation of the lever arm $\alpha$ decreases on hBN compared to SiO$_2$.

Addition energy and plunger gate lever arm are related to the Coulomb peak spacing by $\Delta V_{\rm PG}=E_{\rm add}/\alpha$. It follows, that the peak-spacing fluctuations observed while sweeping $V_{\rm PG}$ over a large range can in principle originate from  (i) fluctuations of single particle level spacing $\Delta$~\cite{pon08, goo13, lib09, kat08a}, (ii) fluctuations of the charging energy $E_{\rm C}$ (i.e. fluctuations in the size of the island), or (iii) fluctuations of the lever arm $\alpha$ (i.e. the position of the charged island).  In graphene quantum dots the single-particle level spacing scales with the number $N$ of electrons in the dot as $\Delta$=$\hbar v_F$/$d \sqrt{N}$~\cite{sch09}. In our measurements, the fluctuations of $N$ are of the order of hundreds, since we measure around 600 subsequent Coulomb peaks for statistical analysis. If we assume that $N$ is the only quantity to vary as $V_{\rm PG}$ is swept, we obtain an upper limit for the standard deviation of the normalized peak-spacing distribution of the order of 0.03,  
 independent of dot size and of the substrate. This is not in agreement with the data of Fig.~\ref{fig03}(e), and we therefore conclude that fluctuations in the quantized level spacing cannot solely account for the experimentally observed distribution of Coulomb peak spacings.

We now turn to the other two possible sources of peak-spacing fluctuations, which are both related to fluctuations of the disorder-induced potential of the graphene nanostructure hosting the QD. In fact, in a rough potential landscape, the dimension of the electron puddle forming the quantum dot and its position  within the etched graphene island might depend in a non-systematic way on the plunger gate voltage $V_{\rm PG}$.   Assuming a simple plate-capacitor model to estimate the charging energy, $E_{\rm C}=e/(4\epsilon_{\rm eff}\epsilon_{0}d)$, and fluctuations in the dot diameter $d$ and the lever arm $\alpha$ to be the main source of variability of the spacing between two Coulomb peaks, we obtain for the standard deviation of the normalized peak spacing distribution  $\sigma \approx \sqrt{\bar{\alpha}^{-2} \sigma_{\alpha}^2 + \bar{d}^{-2} \sigma_{d}^2}$, where $\bar{\alpha}$ ($\bar{d}$)  and  $\sigma_{\alpha}$ ($\sigma_{d}$) are the mean value and the standard deviation of the lever arm (island diameter) fluctuations, respectively.

For graphene QDs on SiO$_{2}$, the standard deviation of the normalized peak-spacing distribution is independent of the nominal size of the dot (i.e. the size of the graphene island) and reads $\sigma^{\rm SiO_{2}}\approx 0.18$, see dashed line in Fig.~\ref{fig03}(e).  Such a value indicates that fluctuations of the dot diameter or of the the lever arm up to 10-20\% can in principle be expected for graphene QDs on SiO$_{2}$. Moreover, the fact that $\sigma^{\rm SiO_{2}}$ does not depends on the geometry of the sample suggests that the potential landscape in the dot is dominated by substrate-induced bulk disorder, while contributions due to edge roughness, which are expected to scale with the size of the sample, play a minor role. 

The situation is opposite for QDs on hBN, where the standard deviation of the normalized peak-spacing distribution shows a clear dependence on the system size.  Assuming a simple model for the $d$-dependance of $\sigma$ (for sufficiently large island size), from the data of Fig.~\ref{fig03}(e)  we obtain $\sigma\approx\sigma^{\rm hBN}+\sigma^{\rm edge}/d\approx 0.01+16/d[\rm nm]$, where $\sigma^{\rm hBN}$ represents the substrate-induced disorder and the second term takes into account the influence of the edges.
These values leads to the conclusion that (i) the substrate induced disorder
in graphene QDs on hBN is reduced by roughly a factor 10 as compared to SiO$_2$, (ii) edge roughness is the dominating source of disorder for QDs with diameters on the order of 100~nm, and (iii) the influence of edge roughness extends for several tenths of nanometers into the bulk. This is in agreement with earlier work on etched graphene nanoribbons on hBN with widths below 80~nm \cite{bis12}, where no significant difference to nanoribbons on SiO$_2$ has been reported. 

In summary, we present an investigation of etched graphene quantum dots of various sizes on hBN. Transport measurements indicate remarkable similarities to QDs on SiO$_2$, but exhibit more stable single-dot characteristics, even at perpendicular magnetic fields up to 9~T. This stability hints at a more homogeneous disorder landscape potential for QDs on hBN as compared to SiO$_{2}$. Further support for this results from a detailed analysis of the peak-spacing distribution of QDs of different sizes realized with identical geometries on both substrates.    
We find that the standard deviation of the peak-spacing distribution shows no size dependence for QDs on SiO$_2$. On the contrary, identical QDs on hBN exhibit a decrease of the standard deviation with increasing dot size. This allows to separate edge from bulk  (i.e. substrate induced) contributions to the disorder potential.  The latter appears to be roughly a factor 10 smaller for devices on hBN as compared to QDs on SiO$_{2}$, so that edge roughness appears to be the dominant source of disorder in QDs on hBN with diameters below 100~nm. The influence of the edges is reduced with increasing device size.
These insights may lead towards cleaner and more controllable graphene QDs.

{Acknowledgment ---}
We acknowledge U. Wichmann for help with the low-noise electronics. 
We thank F. Haupt and J. G\"uttinger for discussions.
Support by the HNF, DFG (SPP-1459 and FOR-912) and ERC are gratefully acknowledged.


\begin{thebibliography}{99}
\bibitem{numberofPeaks}
The exact numbers of Coulomb-peak spacings analyzed for QDs on hBN are: 516 for $d$=110~nm; 658 for $d$=180~nm; 679, 419 and 779 for $d$=300~nm. For  QDs on SiO$_2$: 656 for $d$=110~nm; 592 for $d$=180~nm ; 605, 1003, and 494 for $d$=300~nm.
\end{thebibliography}
\end{document}